\documentclass[
aps,%
12pt,%
final,%
notitlepage,%
oneside,%
onecolumn,%
nobibnotes,%
nofootinbib,
superscriptaddress,%
noshowpacs,%
centertags]%
{revtex4}
\RequirePackage{graphicx}

\def\refitem#1{\relax}

\begin{document}
\title{Beam Energy Dependence of Azimuthal Anisotropy at RHIC-PHENIX }

\author{\firstname{A.} \surname{Taranenko} (for the PHENIX Collaboration)}
\email{arkadij@rcf.rhic.bnl.gov}
\affiliation{Department of Chemistry, Stony Brook University, 
Stony Brook,11779-3400, NY, USA}

\begin{abstract}
Recent PHENIX  measurements of the elliptic ($v_2$) and hexadecapole ($v_4$) 
Fourier flow coefficients for charged hadrons as a function of 
transverse momentum ($p_T$), collision centrality and 
particle species are presented and compared with results from the PHOBOS and 
STAR collaborations respectively. The status of extensions to future PHENIX measurements at  
lower beam energies is also discussed.
\end{abstract}

\maketitle

\section{Introduction}

\begin{figure}[h]
\vspace{-0.2in}
    \centering
        \includegraphics[width=0.98\textwidth]{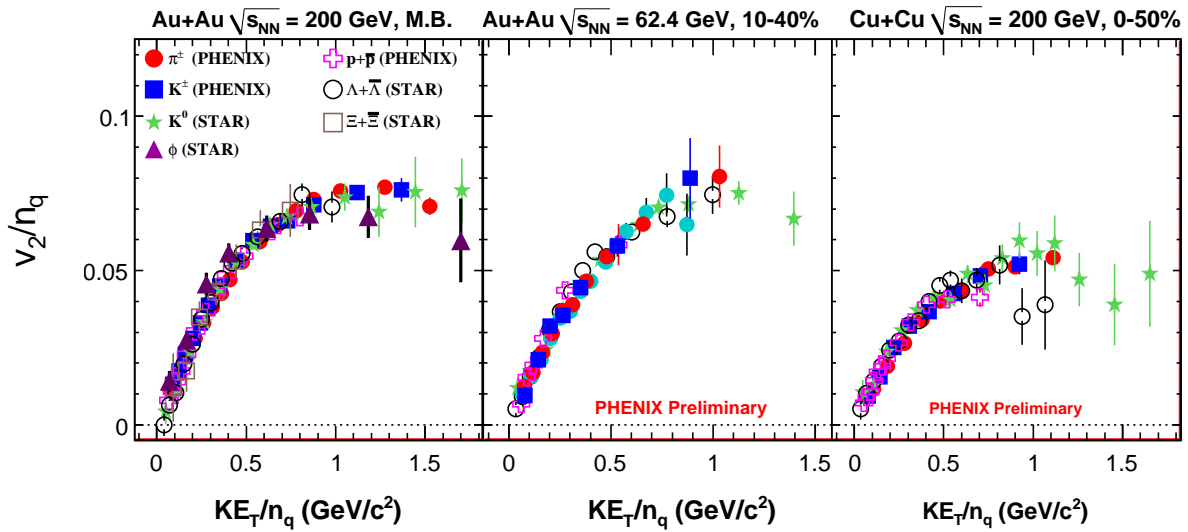}
\vspace{-0.4in}
\caption{$v_2$/$n_q$ as a function of $KE_T$/$n_q$  for identified particle species
obtained in minimum-bias Au+Au collisions at $\sqrt{s_{NN}}$ = 200 GeV (left panel), in 10~-~40$\%$ midcentral 
Au+Au  collisions at $\sqrt{s_{NN}}$ = 62.4 GeV (middle panel) and in  0~-~50$\%$ central 
Cu+Cu collisions at $\sqrt{s_{NN}}$ = 200 GeV (right panel).}
\label{v2star2phenix}
\end{figure}

The discovery of large  elliptic flow ($v_{2}$) for all particle species  
studied at the Relativistic Heavy Ion Collider (RHIC), is a key piece of evidence for the 
creation of hot and dense partonic matter in ultra relativistic nucleus-nucleus collisions 
\cite{Adcox:2004mh,Voloshin:2008dg}. This is well supported by the observed agreement between 
differential flow measurements and 
calculations that model an essentially locally equilibrated quark gluon plasma (QGP) having  
a small value for the specific shear viscosity ($\eta/s$) \cite{Romatschke:2007mq,Lacey:2006bc,Xu:2007jv,Heinz:2009cv}. 
After ten years of experiments at RHIC, the most extensive set of flow measurements 
for various hadrons with different masses,
charges, quark content and hadronic cross-sections became available for the first time
in the history of heavy-ion collisions. They show that, for a given centrality, elliptic 
flow for all observed hadrons [at RHIC] scale to a single curve when plotted 
as $v_2/n_q$ versus $KE_{T}/n_q$, where $n_q$ is the number of constituent quarks in a given 
hadron species and $KE_{T}$ is the transverse kinetic  energy for these hadrons \cite{Adare:2006ti,Lacey:2006pn}.  
Such scaling is illustrated in Fig.~\ref{v2star2phenix} where a compilation of elliptic 
flow results for identified hadrons, measured by STAR \cite{Adams:2004bi,Abelev:2007qg,:2008ed,Shi:2009jg}
and PHENIX \cite {Adare:2006ti,Taranenko:2007gb,Issah:2008rk,Shimomura:2009hb} for minimum-bias Au+Au collisions at $\sqrt{s_{NN}}$ = 200 GeV (left panel), 10-40$\%$ midcentral 
Au+Au  collisions at $\sqrt{s_{NN}}$ = 62.4 GeV (middle panel) and 0-50$\%$ central 
Cu+Cu collisions at $\sqrt{s_{NN}}$ = 200 GeV (right panel). 
The observation that all of these data show scaling indicates that the agreement between the 
STAR and PHENIX $v_2$ data is better than $\sim 15$\%, otherwise, the scaling would be broken. 
More importantly, the observed scaling suggests that the bulk of the elliptic flow at RHIC 
energies is partonic, rather than hadronic \cite{Adare:2006ti,Lacey:2006pn,Afanasiev:2007tv}.

Lowering the collision energy and studying the energy dependence of anisotropic flow 
allows a search for the onset of the transition to a phase with partonic degrees of 
freedom at an early stage 
of the collision \cite{Lacey:2006bc,Aggarwal:2010cw}. 
In the present work, we review the most recent flow measurements performed by the PHENIX 
collaboration at top RHIC energies, and give the most comprehensive comparison of these 
results with measurements from both  PHOBOS and STAR collaborations.
We also discuss the progress of our ongoing flow analysis at lower collision energies at RHIC.

\section{Results}
Future progress in the extraction of  the transport properties of hot and dense partonic
matter from flow measurements at RHIC, depends strongly on further developments of theoretical 
models, as well as a better understanding of the systematic errors associated with 
flow measurements. 
First comparisons of differential  elliptic flow data from the top RHIC 
energy ($\sqrt{s_{NN}}$ = 200 GeV) with viscous relativistic 
hydrodynamic simulations, demonstrate that a 20\% uncertainty in the measured $v_2$ leads to  
$\simeq$ 60-70\% unsertainty in the extracted value of specific shear viscosity (~$\eta/s$~) 
\cite{Romatschke:2007mq,Heinz:2009cv,Lacey:2010fe}. The role of fluctuations and so-called ``nonflow'' 
correlations can be important for such measurements \cite{Ollitrault:2009ie}. 
The next lesson is that the simultaneous measurement of $v_2$ and higher harmonics 
such as hexadecapole flow ($v_4$) and triangular flow ($v_3$), 
can help to better constrain $\eta/s$ \cite{Luzum:2010ae,Alver:2010dn,Schenke:2010rr}.

PHENIX has addressed many of these issues via a new set of measurements of charged 
hadron $v_2$ and $v_4$ \cite{Adare:2010ux}.
These measurements were performed in the 
two PHENIX central arms ($\left| \eta \right| \le 0.35$) relative to event 
planes obtained from four separate reaction-plane detectors in the range $1.0 
< \left| \eta \right| < 3.9$, see left panel of Fig.~2. 
Multiple event planes allow a search for 
possible $\Delta\eta$-dependent non-flow contributions that would influence the 
magnitude of $v_{2,4}$, which is crucial for reliable extraction of 
transport coefficients.

\begin{figure}[ht]
\vspace{0.14in}
\begin{center}
\resizebox{
\textwidth}{!}{
\resizebox*{10cm}{8.57cm}{
\includegraphics{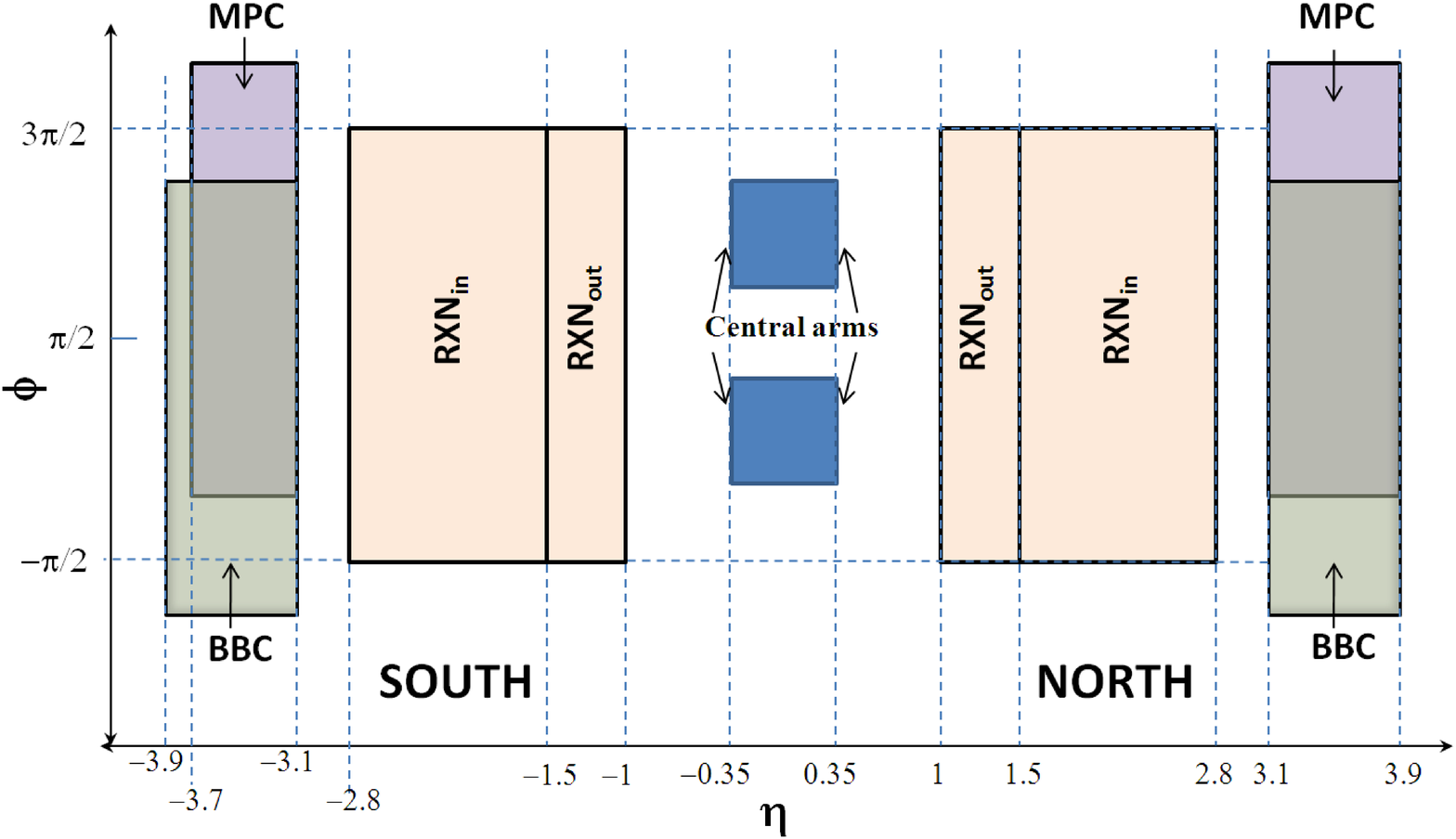}}
\resizebox*{10cm}{8.5cm}{
\includegraphics{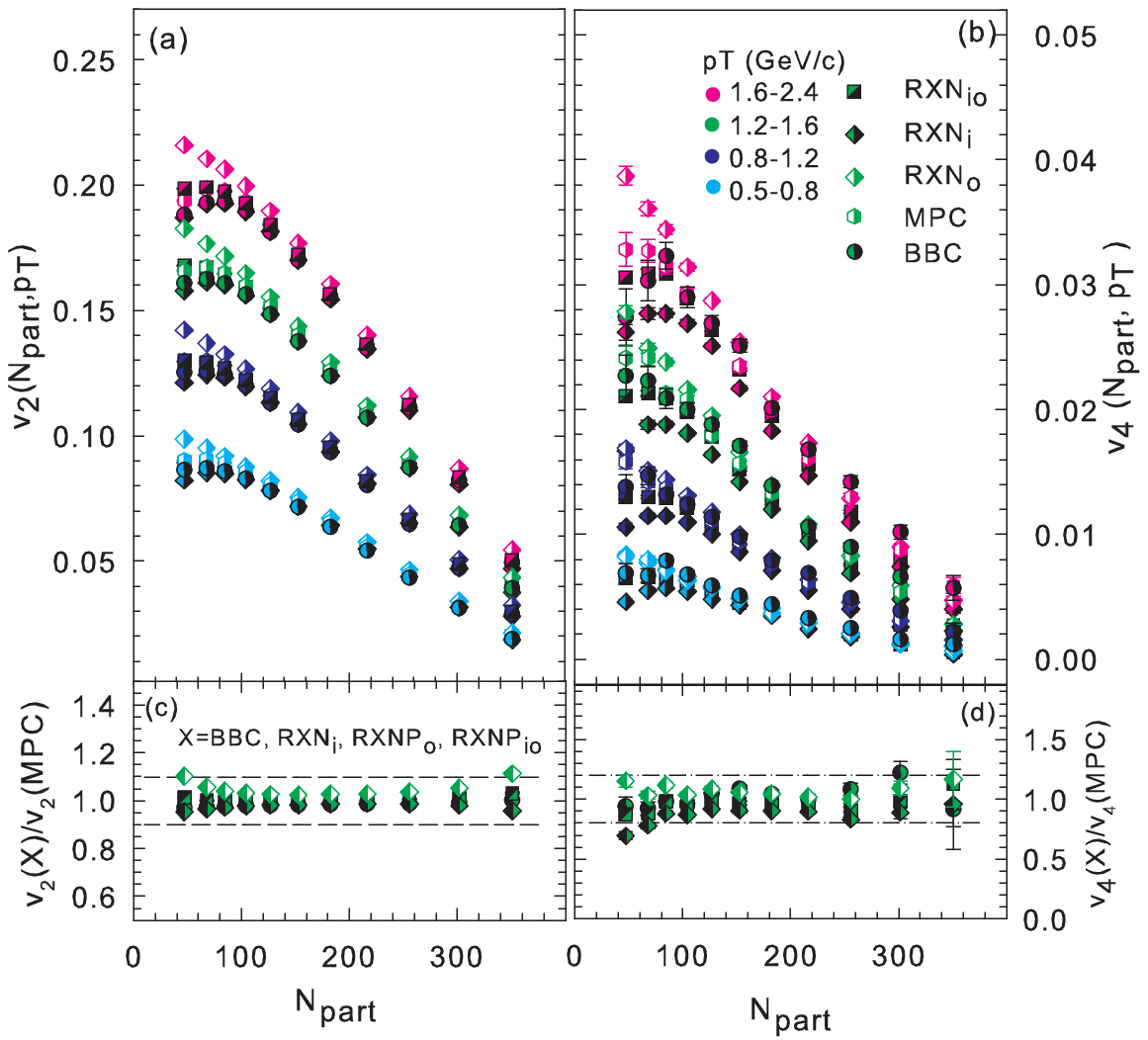}}}
\caption{ Left panel: The  azimuthal angle $\phi$ and pseudo-rapidity $\eta$ acceptance of 
PHENIX detectors used for event plane reconstruction, together with the central arm acceptances for 
charged hadron measurements. Right panel: comparison of $v_2$ vs.  $N_{\rm part}$ and $v_4$ 
vs.  $N_{\rm part}$ for charged hadrons obtained with several reaction 
plane detectors for the $p_T$ selections indicated \cite{Adare:2010ux}. Ratios for the $p_T$ range 
1.2--1.6 GeV/$c$ are shown in (c) and (d). }
\end{center}
\vspace{-0.5cm}
\label{rpdet2}
\end{figure}

The right panel of Fig. 2  compares the double differential flow 
coefficients $v_{2,4}(p_T, N_{\rm part})$ for event-plane detectors 
spanning the range $1.0 < \left| \eta \right| < 3.9$.  Within systematic errors, they 
agree to better than $\sim$ 5\% (10\%) for $v_2$ ($v_4$) in mid-central 
collisions and approximately 10\% (20\%) in central and peripheral events. This 
agreement indicates that the present  measurements of $v_2$ and $v_4$
are not affected by  $\Delta\eta$- and $p_T$-dependent non-flow contributions 
for $p_T \alt 3$ GeV/$c$ and centrality $\le 60\%$ \cite{Adare:2010ux}.

\begin{figure}[!htb]
\begin{tabular}{cc}
\vspace{0.2in}
\includegraphics[width=0.57\linewidth]{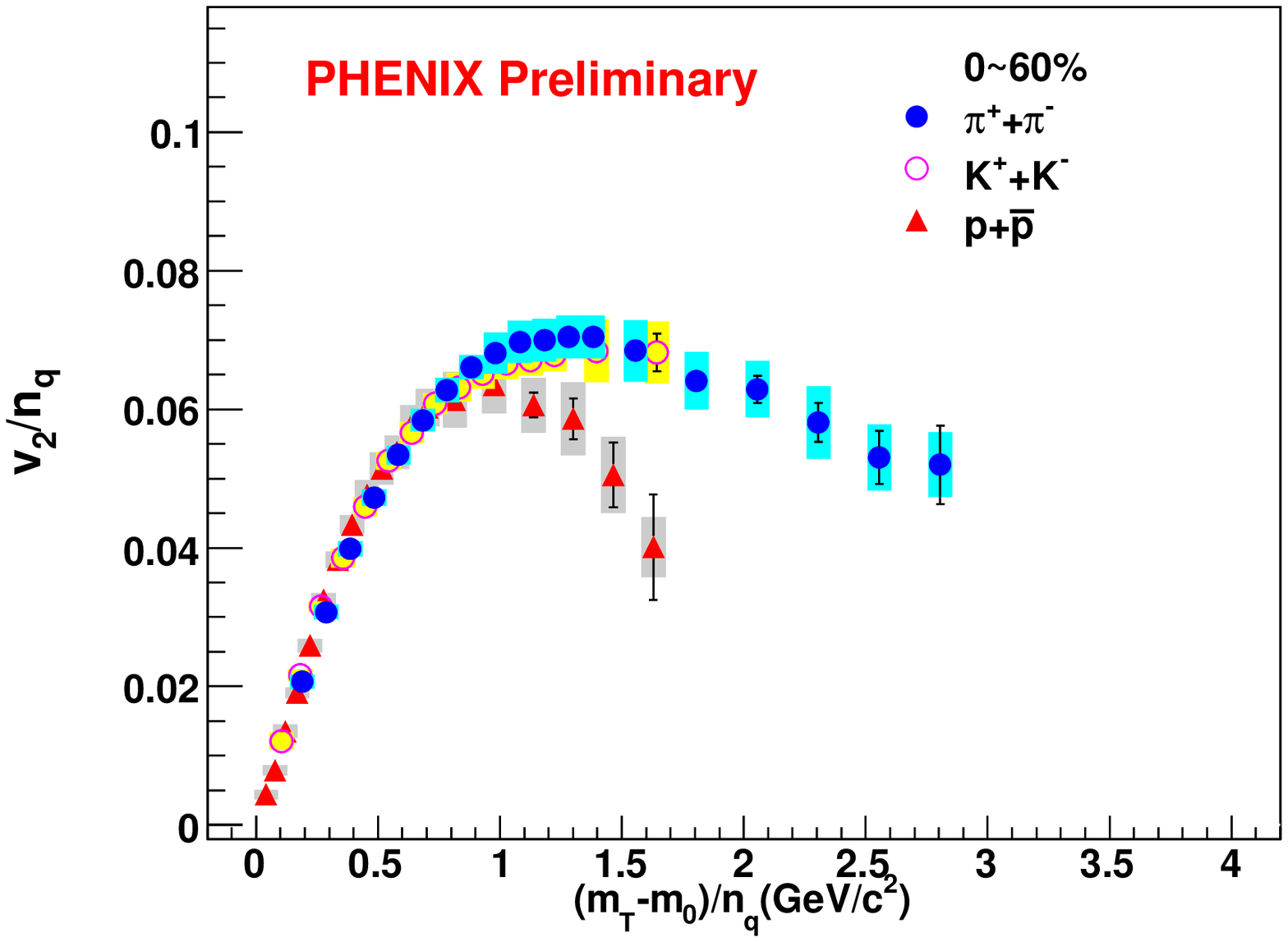} &
\includegraphics[width=0.45\linewidth]{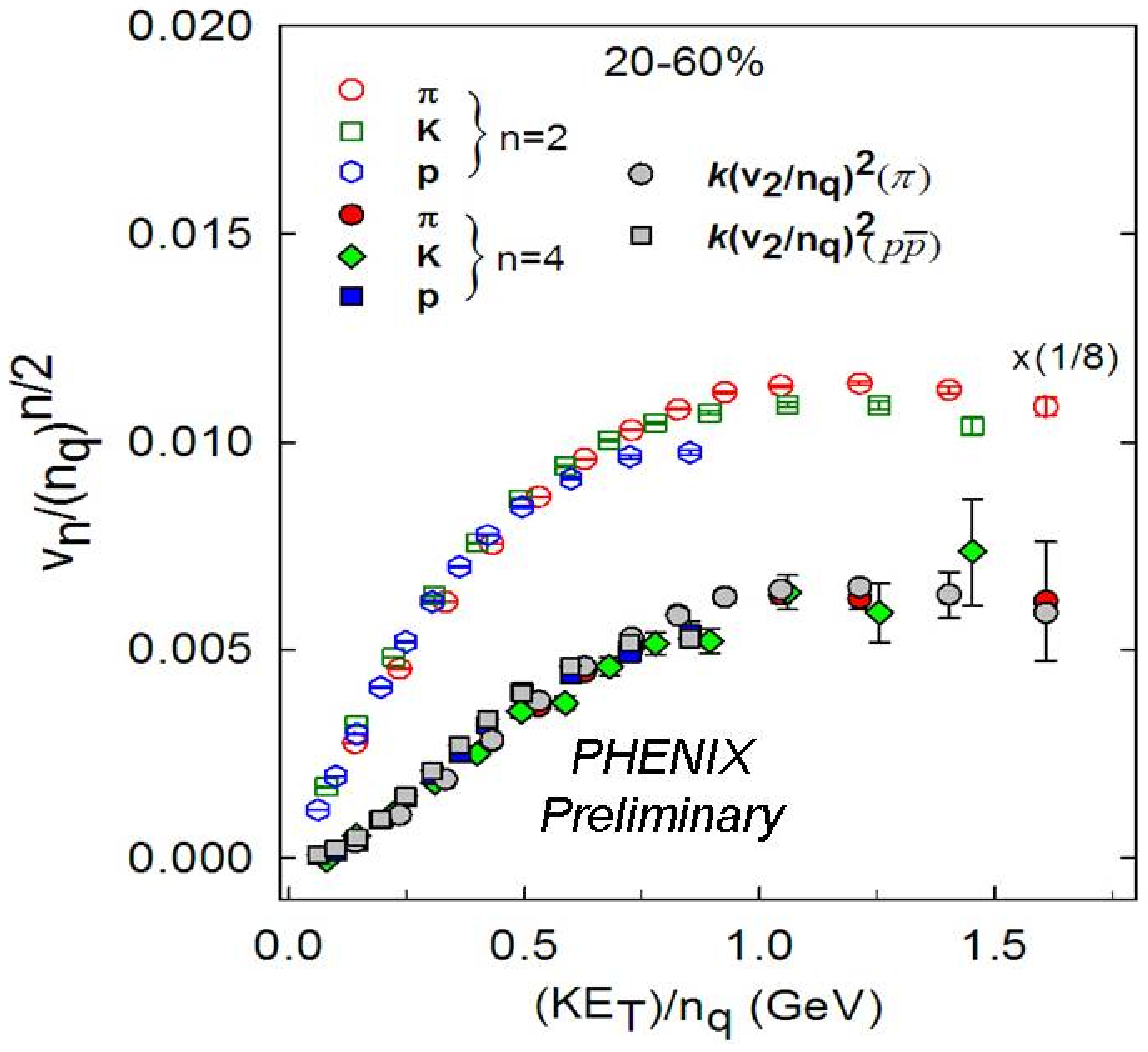} \\
\end{tabular}
\vspace{-0.2in}
\caption{$v_2$/$n_q$ as a function of $KE_T$/$n_q$  for identified charged hadrons obtianed 
in 0-60\% central Au+Au collisions at $\sqrt{s_{NN}}$ = 200 GeV [~left panel~] and 
$v_n$/$n_q^{n/2}$ as a function of $KE_T$/$n_q$ for the same particle species [~right panel~].}
\label{v2v4ket}
\end{figure}

The recently installed time-of-flight detector (TOFw) with intrinsic timing resolution of $\simeq$ 75 ps significantly
improved the particle identification capabilities, as well as the acceptance of PHENIX for flow measurements of identified
hadrons. Time-of-flight measurements in conjunction with measured momentum and flight-path length allow proton/kaon
separation up to p$_T\simeq$ 4.5 GeV/c via mass-squared calculations. 
With the combined measurements of mass-squared and photon yield from the Aerogel Cherenkov counter (ACC), one can extend
the $v_2$ measurements of charged pions and protons up to $p_T\simeq$ 6 GeV/c \cite{Huang:2009zzh}. This is shown in the left panel 
of Fig.~\ref{v2v4ket} where measured $v_2$ values for charged pions , kaons and protons are plotted 
in the scaled variables $v_2/n_q$ versus
$KE_{T}/n_q$. These scaled data indicate that the universal $KE_{T}/n_q$ scaling, evident at 
low $KE_{T}$, is broken after $KE_{T}/n_q \agt 1GeV/c^2$, indicating a possible change in the 
physics.

\begin{figure}[!htb]
\begin{tabular}{cc}
\vspace{0.25in}
\includegraphics[width=0.5\linewidth]{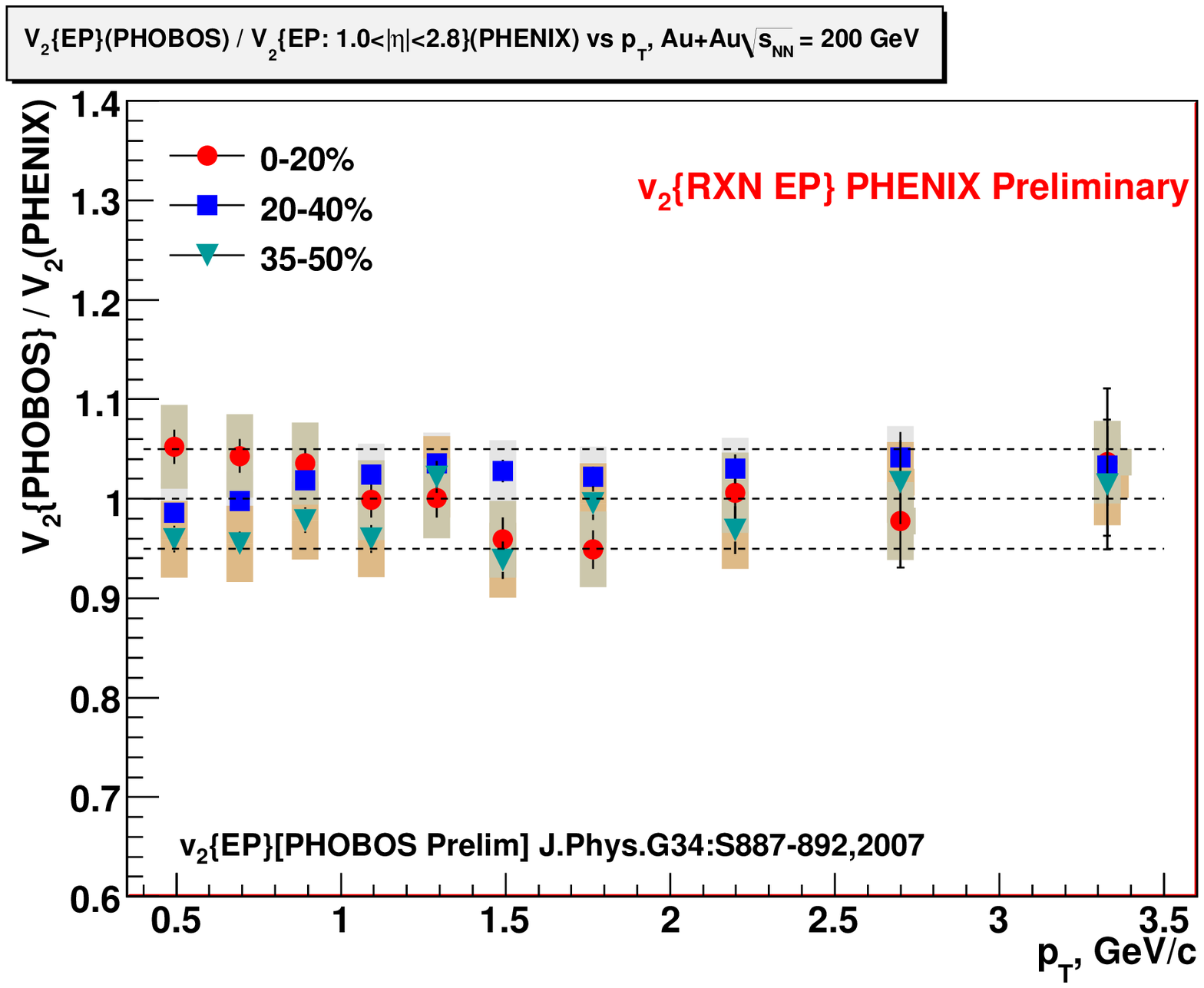} &
\includegraphics[width=0.5\linewidth]{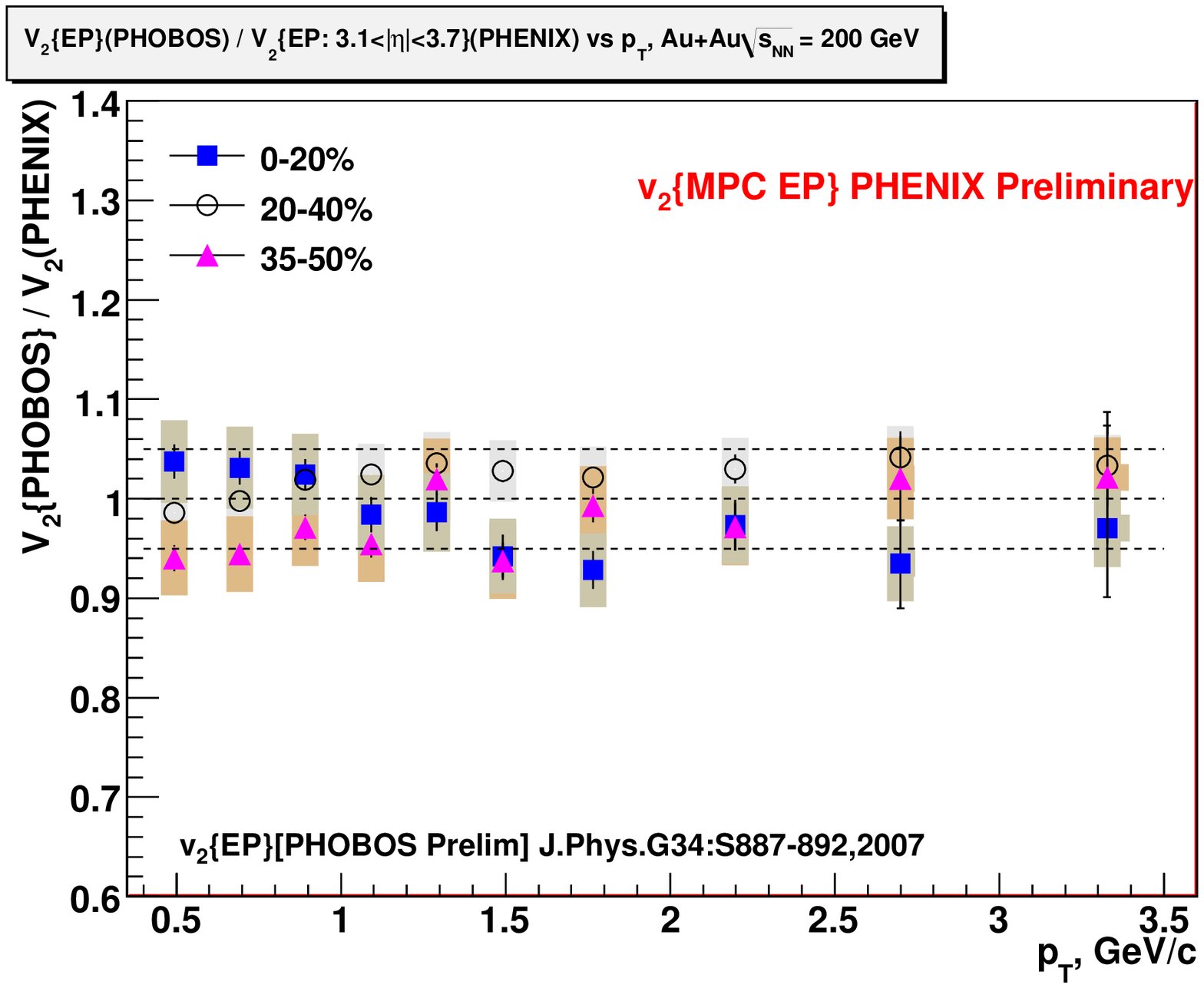} \\
\end{tabular}
\vspace{-0.35in}
\caption{$p_T$ dependence of the ratio of  PHOBOS $v_2$ values \cite{Alver:2007rs}  obtained by 
event plane method $v_2$\{EP\} ($2.05<|\eta|<3.2$)
to the PHENIX $v_2$ measured using 
 RXN ($1.0<|\eta|<2.8$) [left panel ] and MPC ($3.1<|\eta|<3.7$) [right panel ]. }
\label{v2phobos}
\end{figure}

Statistically significant measurements of $v_4$ for identified hadrons also enable a 
study of a related scaling relation between $v_4$ for different particle species, as 
well as the ratio $v_4/v_2^2$. The right panel of Fig.~\ref{v2v4ket} shows the result of such 
a study in which both $v_2$ and $v_4$ are plotted using the generalized scaling 
variables, $v_n$/$n_q^{n/2}$ vs. $KE_T$/$n_q$.
Further detailed studies of this generalized scaling for different flow harmonics, as well as 
the $KE_{T}/n_q$ range where scaling breaks down, should help to better understand the 
transition from soft to hard physics \cite{Lacey:2010fe} and the process of hadronization at RHIC.\\
An important step toward reducing the systematic  uncertainty associated with flow measurements at RHIC, is a detailed 
comparison of differential flow results obtained by different collaborations - PHOBOS, PHENIX and STAR. 
One way to show such comparisons is to plot the respective ratios of these measurements as a function of $p_T$ and centrality, 
while ensuring that both centrality and mean $p_T$ are the same. For reference, we have used 
PHENIX flow measurements of charged hadrons with respect to the MPC ($3.1<|\eta|<3.7$) and RXN 
combined  ($1.0<|\eta|<2.8$) event planes \cite{Adare:2010ux} from recent data obtained in 2007.
The ratios  $v_2$(PHOBOS)/$v_2$(PHENIX) are plotted in Fig ~\ref{v2phobos}
as a function of $p_T$ for several bins in collision centrality as indicated.
 The values for PHOBOS $v_2(p_T)$ were obtained  
for charged hadrons with $0<\eta<1.5$ with respect to the second order event plane measured
at $2.05<\eta<3.2$ \cite{Alver:2007rs}.

\begin{figure}[!htb]
\begin{tabular}{cc}
\vspace{-0.1in}
\includegraphics[width=0.5\linewidth]{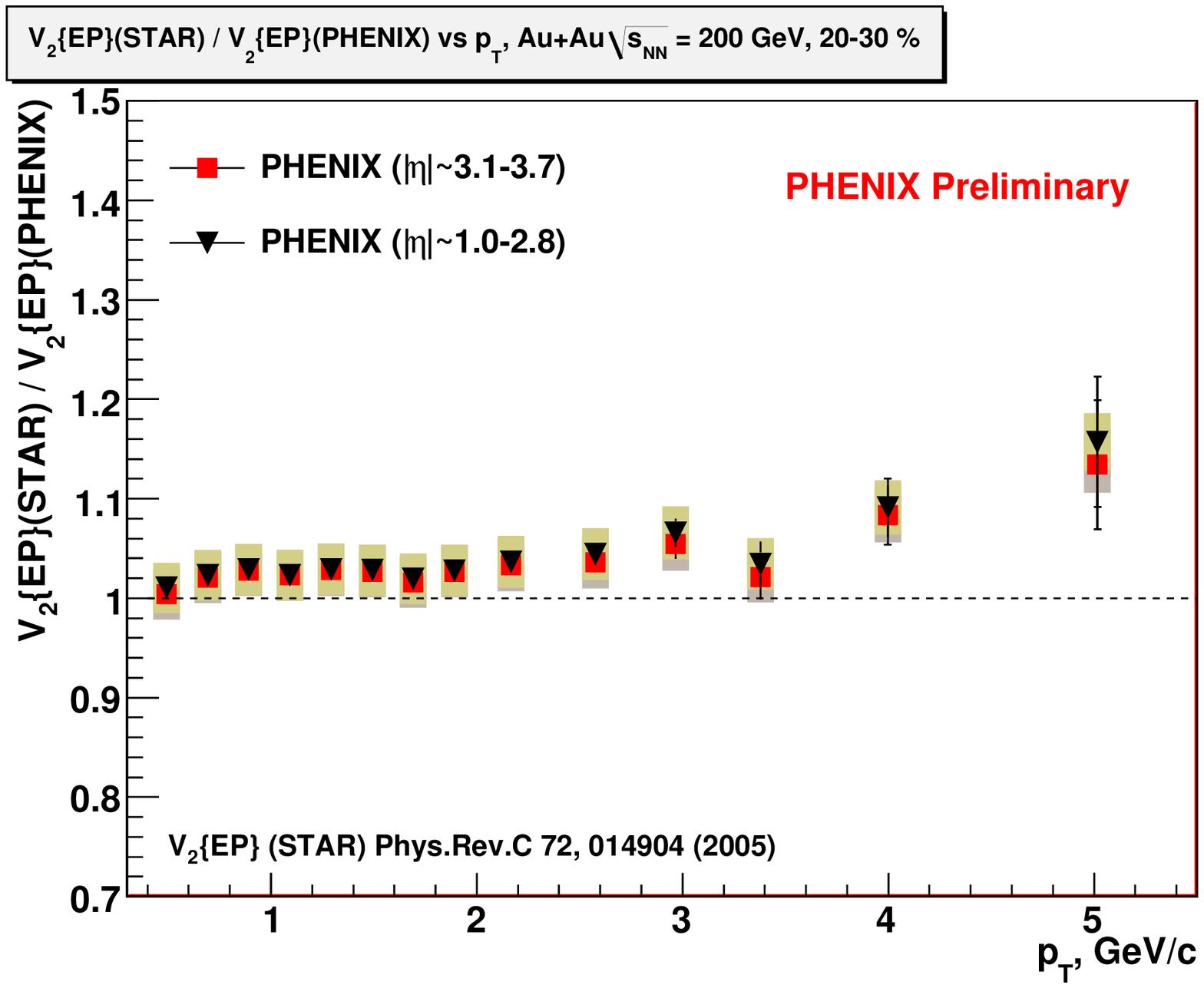} &
\includegraphics[width=0.5\linewidth]{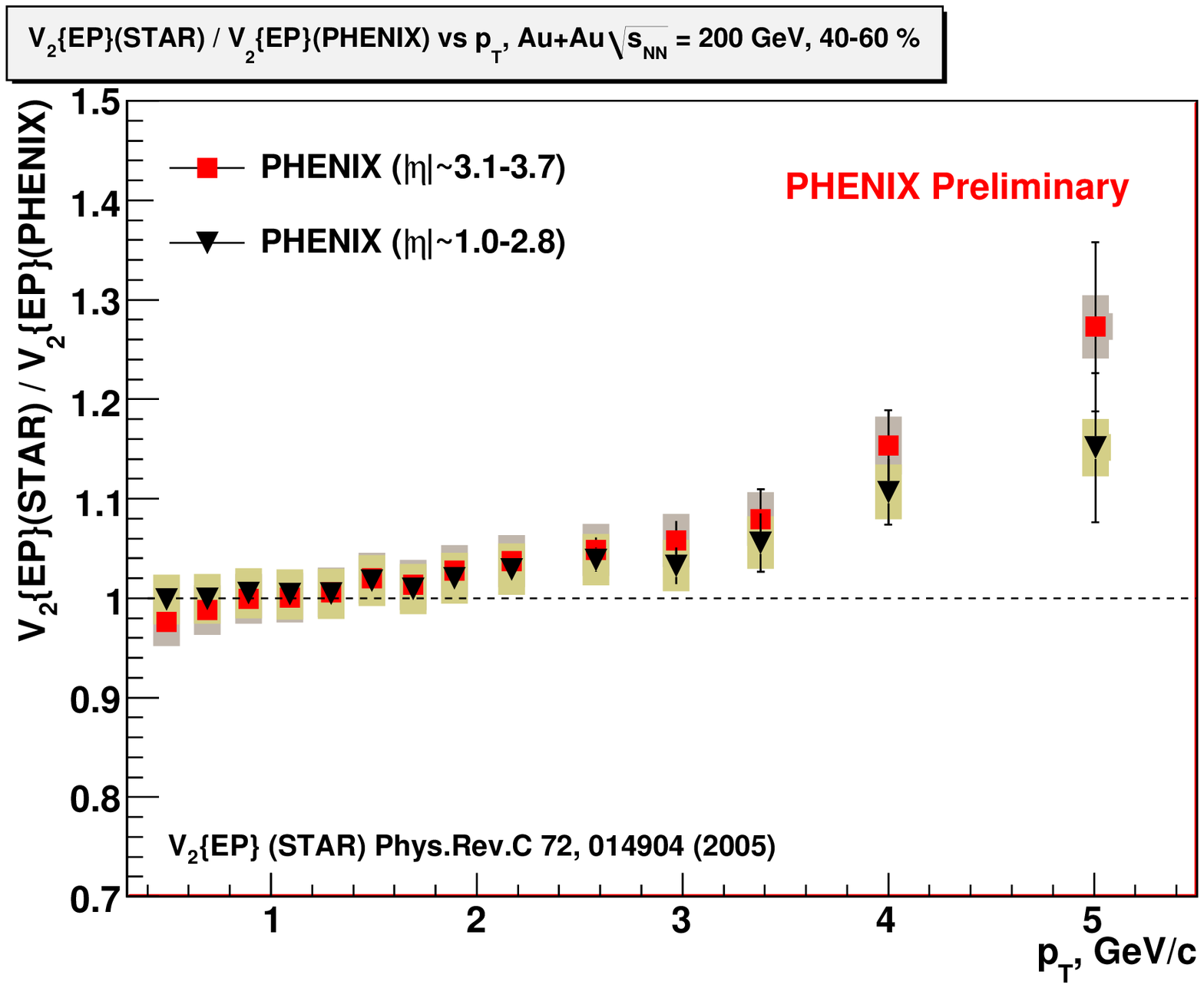} \\
\includegraphics[width=0.5\linewidth]{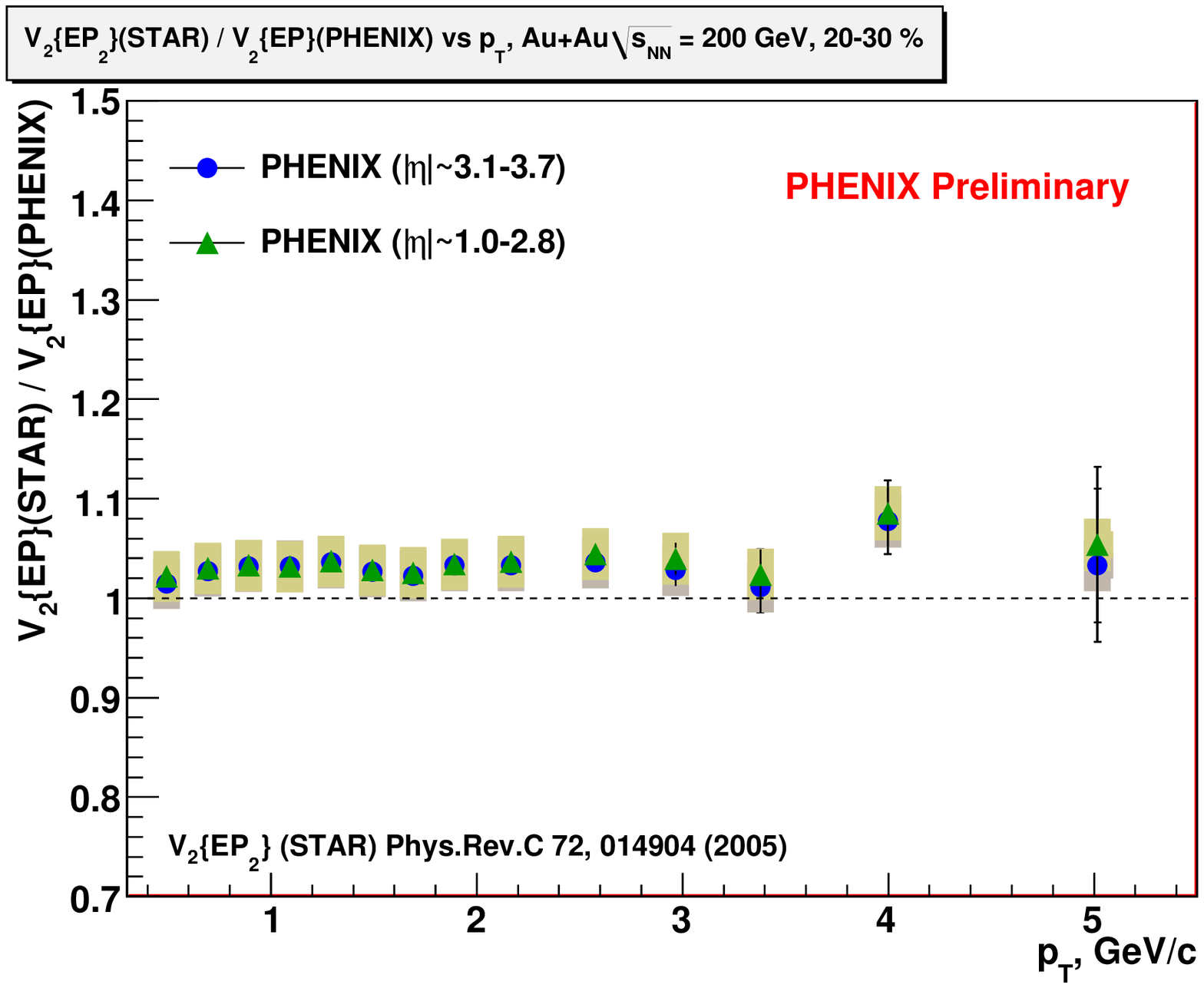} &
\includegraphics[width=0.5\linewidth]{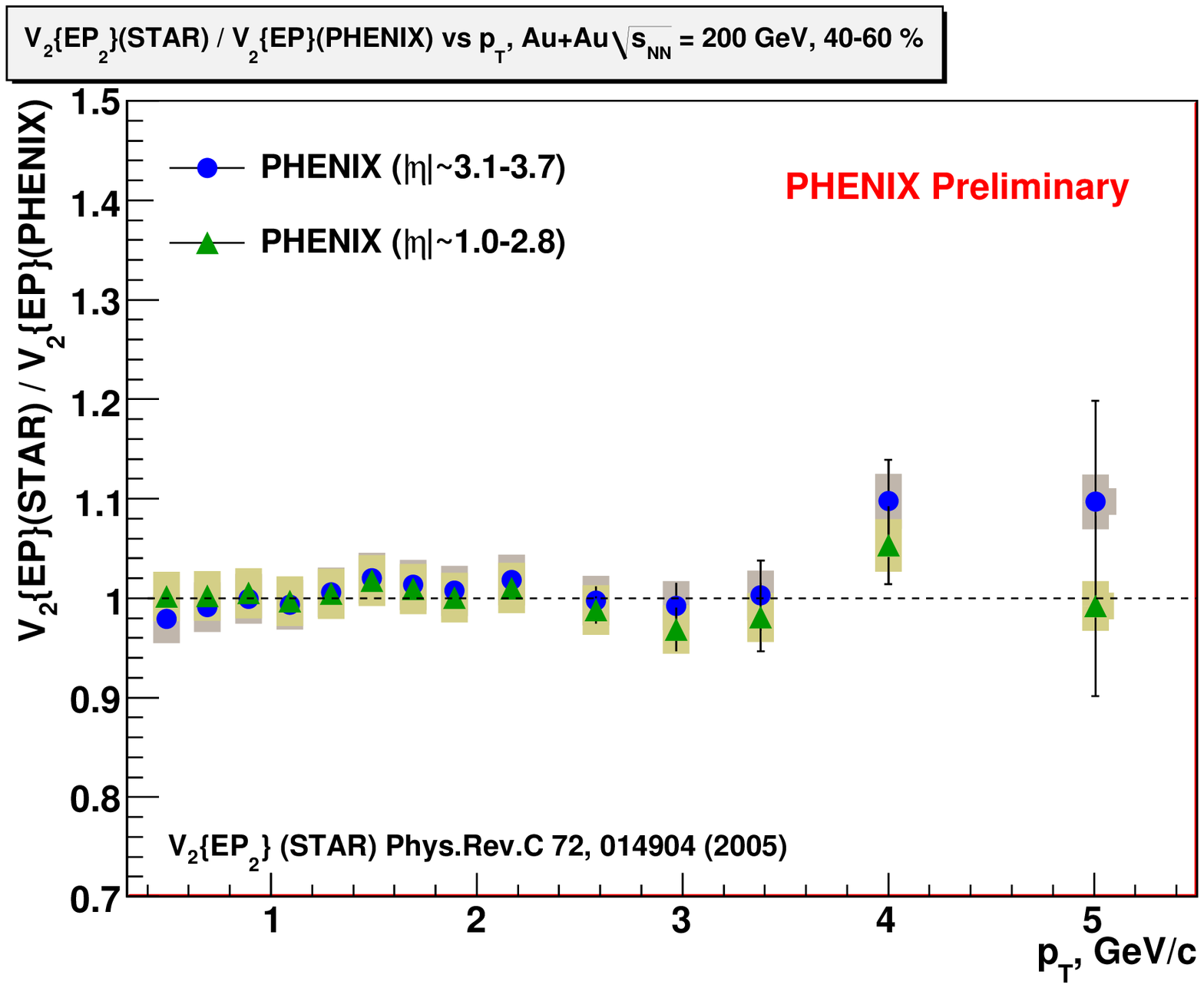} \\
\end{tabular}
\vspace{-0.3in}
\caption{$p_T$ dependence of the ratio of  STAR $v_2$ values for charged hadrons obtained by 
standard [upper panel ] and modified [ lower panel ] event plane method $v_2$\{EP\} using central TPC ($|\eta|<1.0$)
to the PHENIX $v_2$ values obtained by event plane method using MPC/RXN event planes
for 30-40\% and 40-60\% midcentral Au+Au collisions.}
\label{v2star1}
\end{figure}

The ratio plots indicate that the difference between PHENIX/PHOBOS $v_2$ results are $\le 5\%$ 
and do not show a significant $p_T$ dependence.
The STAR $v_2$ for charged hadrons from Au+Au collisions at $\sqrt{s_{NN}}$~=~200 GeV measured with 
respect to the event plane from the FTPC detector ($2.5<|\eta|<4.0$) are not available. Therefore our  
comparison was made with STAR results obtained via the standard event plane method using 
the central TPC ($0<|\eta|<1$) $v_2$\{EP\}, and a modified event plane method where 
particles within $|\Delta\eta| < 0.5$ around the highest $p_T$ 
particle were excluded for the determination of the modified 
event plane $v_2$\{$EP_2$\}; this procedure reduces some of the non-flow effects at 
high $p_T$ \cite{Adams:2004bi,:2008ed}.  The ratios $v_2$(STAR EP)/$v_2$(PHENIX) are plotted in the top panel 
of Fig ~\ref{v2star1} for 30-40\% and 40-60\% midcentral Au+Au collisions at $\sqrt{s_{NN}}$~=~200 GeV.

\begin{figure}[!htb]
\begin{tabular}{cc}
\vspace{0.3in}
\includegraphics[width=0.55\linewidth]{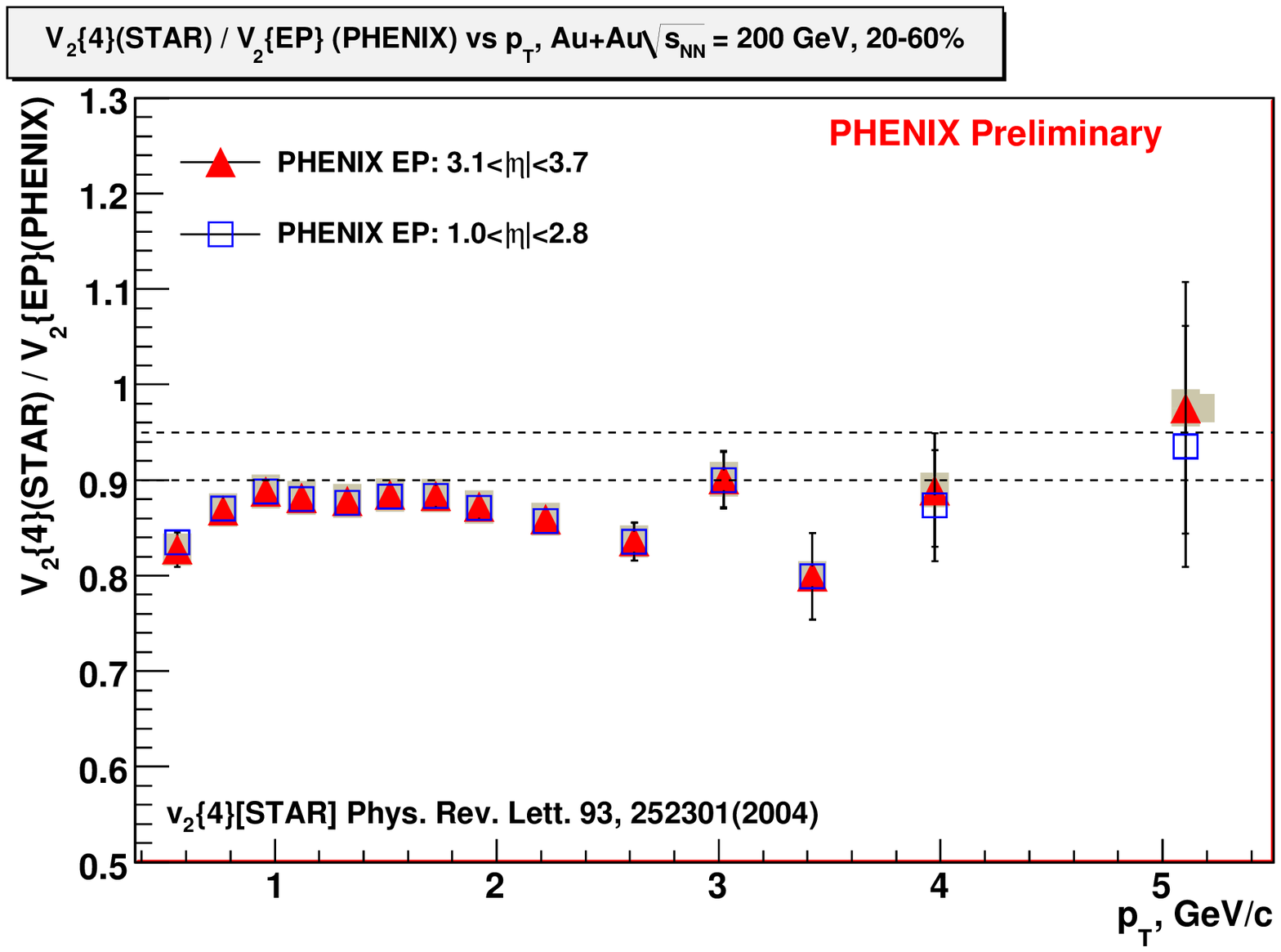} &
\includegraphics[width=0.5\linewidth]{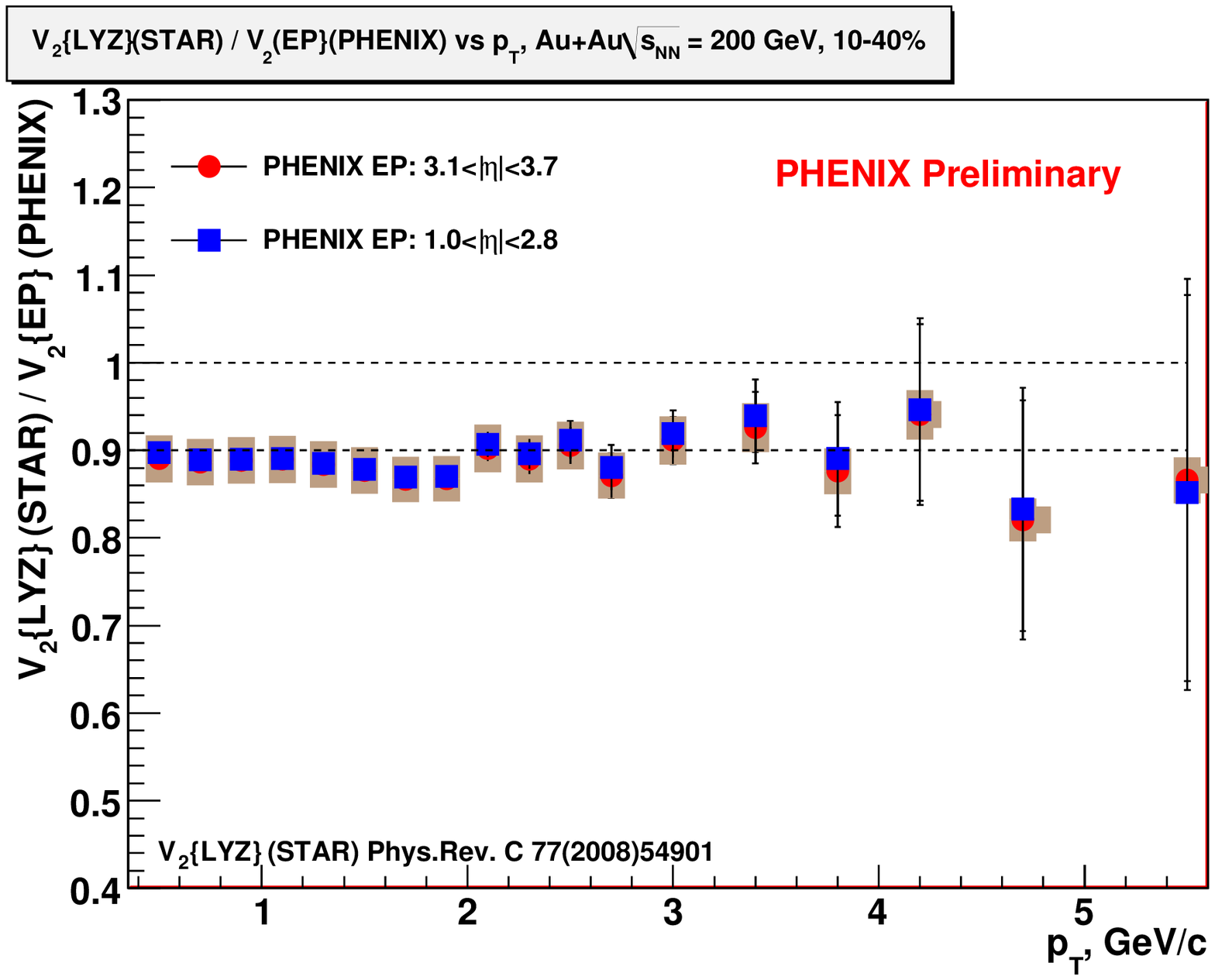} \\
\end{tabular}
\vspace{-0.2in}
\caption{$p_T$ dependence of the ratio of STAR $v_2$ values for charged hadrons obtained by four-particle
cumulant method  $v_2$\{4\} [left panel] and Lee-Yang Zero method [right panel]
using charged tracks from central TPC ($|\eta|<1.0)$ to the PHENIX $v_2$ values obtained by event plane method using MPC/RXN event planes.}
\label{v2star2}
\end{figure}

The difference between STAR/PHENIX $v_2$ values 
is less than 2-5\% below $p_T \simeq$2.5 GeV/c.  At higher
transverse momentum, STAR $v_2$ is systematically larger than the  PHENIX $v_2$ and the ratio tends to grow with 
p$_T$ reaching the value of  20\% at $p_T \simeq $5.5 GeV/c. 
The difference in $v_2$ values at higher $p_T$ can be attributed to non-flow effects due to di-jets which are mostly
suppressed by the rapidity gaps in the  case of the PHENIX measurements. 
The lower panels of Figure ~\ref{v2star1} show the same 
ratios but in this case the STAR $v_2$ data were obtained by the modified  event plane method $v_2$\{EP$_{2}$\}. 
Here, the difference between STAR/PHENIX ratios at high p$_T$, are much smaller, but still persist 
on the level of 5-10\% at p$_T\simeq$ 5.5 GeV/c for mid-central collisions. The comparison for other centralities
can be found elsewhere \cite{bnl09}.
Figure ~\ref{v2star2} show the comparison of the same PHENIX $v_2$ data set with STAR $v_2$ 
results obtained using multi-particle methods. That is, the four particle cumulant method (left panel) 
and the Lee-Yang Zero method (right panel).

The ratio plots indicate that
STAR $v_2$ results obtained using multi-particle methods are smaller than PHENIX $v_2$\{EP\} event plane results
by 10-12\%. Note however that this difference does not show a p$_T$ dependence and can be attributed 
to the expected difference in the eccentricity fluctuations for the two- and multi-particle measurements.

\section{Collision energy dependence of azimuthal anisotropy}
\begin{figure}[ht]
\vspace{0.14in}
\begin{center}
\resizebox{
\textwidth}{!}{
\resizebox*{10cm}{8.7cm}{
\includegraphics{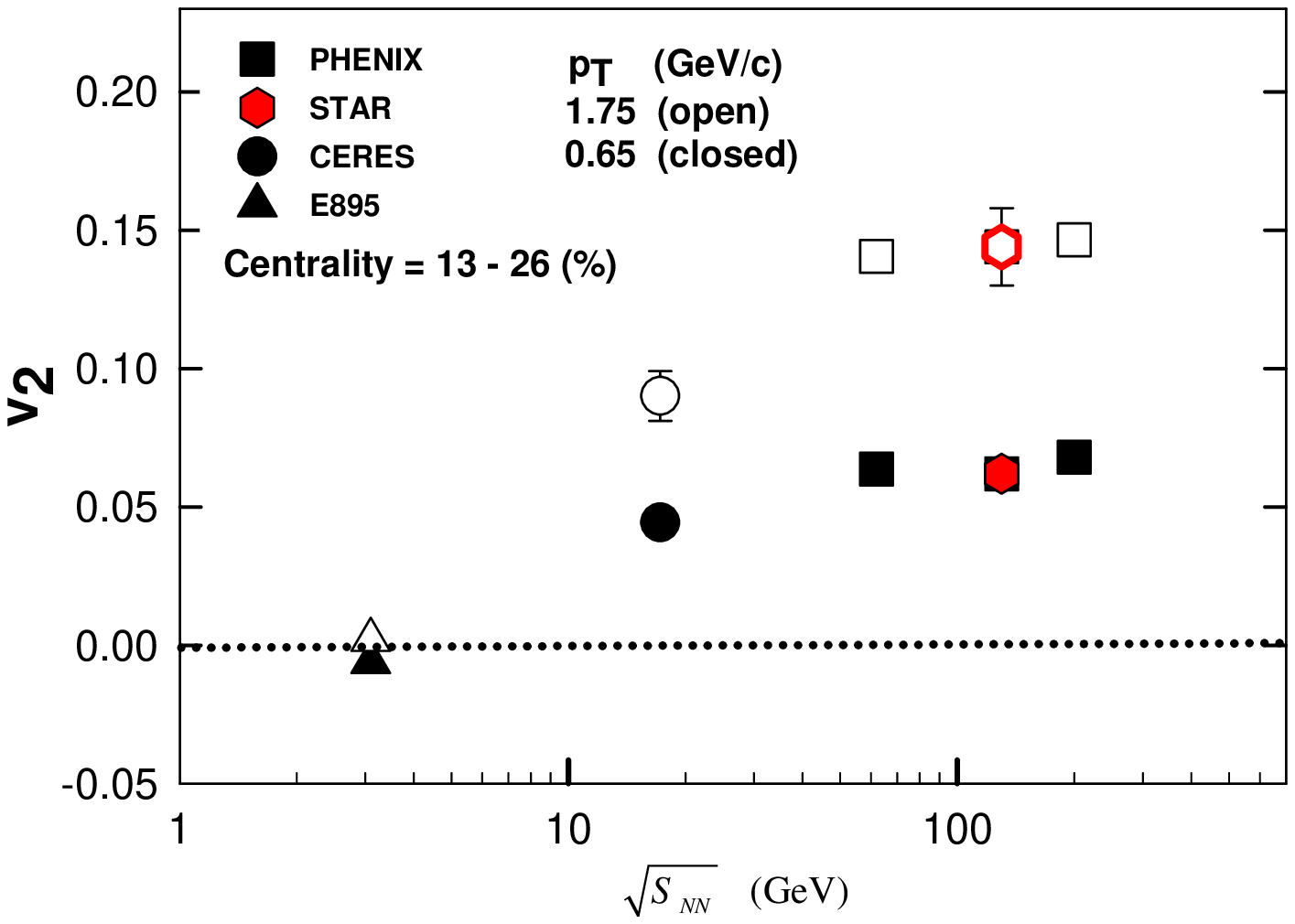}}
\resizebox*{10cm}{8.5cm}{
\includegraphics{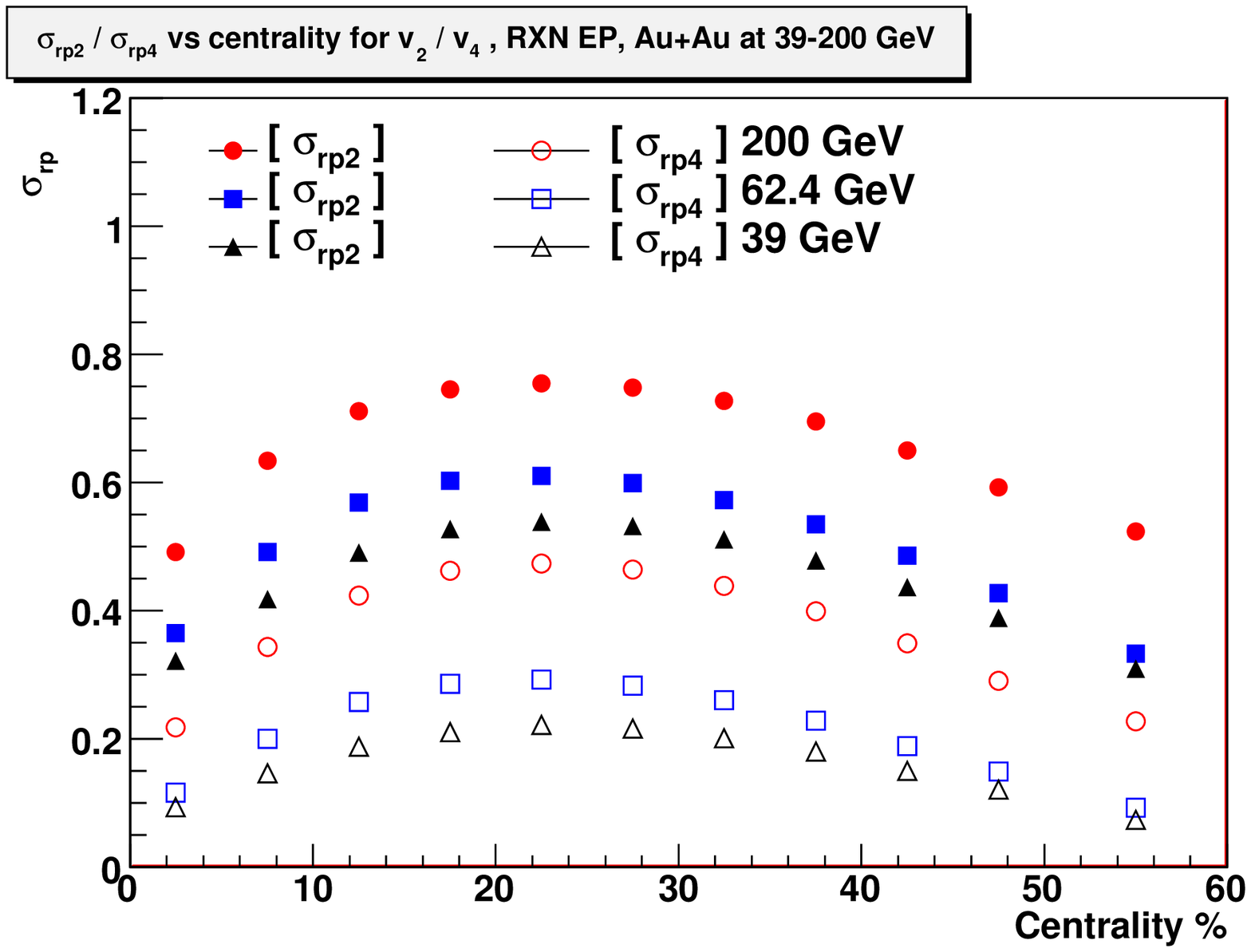}}}
\caption{ Left panel: differential $v_2$ vs. $\sqrt{s_{NN}}$  for charged  hadrons in nucleus-nucleus
collisions for centrality cut of 13-26\% and two different mean $p_T$ values. From \cite{Adler:2004cj}. Right panel: centrality dependence of event plane resolution factors for $v_2$ (~closed symbols~) and $v_4$ (~closed symbols~) measurements using PHENIX RXN detector for Au+Au 
collisions at $\sqrt{s_{NN}}$ = 39~-~200 GeV. }
\end{center}
\vspace{-0.5cm}
\label{v2beam}
\end{figure}

A large number of elliptic flow measurements have been performed by many experimental groups 
at SIS, AGS, SPS and RHIC energies over the last twenty years. However, the fact that these 
data were not obtained under the same experimental conditions, do not allow a detailed and 
meaningful comparison in most cases -- the situation at RHIC is of course somewhat better.
Experimental differences include: (a) different centrality selection, (b) 
different transverse momentum acceptance, (c) different particle species, (d) different rapidity coverage and
(e) different methods for flow analysis.  The results from PHENIX and STAR indicate that the 
magnitudes and trends of the differential elliptic flow , $v_2$(~$p_T$, centrality~), 
change very little over the collision energy range
$\sqrt{s_{NN}}$~=~62~-~200~GeV, indicating saturation of the excitation function
for $v_2$ at these energies \cite{Adler:2004cj,Abelev:2007qg}, see left panel of Fig.~\ref{v2beam}.

The figure also shows that the differential elliptic flow for charged hadrons increase by almost 50\% 
from the top SPS energy of 17.3 GeV to $\sqrt{s_{NN}}$~=~62 GeV at RHIC. This conclusion 
is based on the comparison of PHENIX results 
with published results from CERES collaboration \cite{Agakichiev:2003gg}. However, the comparison of STAR 
results for  $v_2$(~$p_T~$) for charged pions and protons at $\sqrt{s_{NN}}$~=~62 GeV 
with results from NA49 Collaboration  at 17.3 GeV \cite{Alt:2003ab}, leads to the conclusion that the differential $v_2$ change
only by 10-15\% from top SPS energy to RHIC \cite{Abelev:2007qg}. This may indicate that the existing  flow 
results at the SPS are prone to large systematic uncertainties which are not yet well understood. 
Given the fact that the energy density  
increases by approximately 30\% over the range $\sqrt{s_{NN}}$~=~62.4~-~200 GeV,  
the apparent saturation of differential $v_2$ at RHIC could be an indication of a softening  
of the equation of state due to the crossover transition. 

In Run 2010 the PHENIX Collaboration collected ~$\sim 5 \times 10^8$ minimum-bias 
Au+Au events at $\sqrt{s_{NN}}=$ 62.4 GeV and  ~$\sim 2 \times 10^8$ minimum-bias Au+Au events at $\sqrt{s_{NN}}=$ 39 GeV.
The analysis of these data is progressing very well. The right 
panel of Fig.~\ref{v2beam} shows the centrality dependence of the event plane resolution 
factors for $v_2$ and $v_4$ measurements obtained for the second order event planes from the RXN 
detector for the collision energies 39, 62.4 and 200 GeV. They indicate that reliable extraction 
of the flow harmonics should be straightforward. Note that this large data set in conjunction with 
an improved event plane resolution, gives an equivalent of $\simeq$ 30-fold increase in statistics 
over the previous measurement of $v_2$ at 62 GeV \cite{Adler:2004cj}. We expect that they will allow 
detailed measurements of the differential $v_2$ and $v_4$ as a function of $p_T$ and  centrality  
for several particle species at both energies.

\begin{acknowledgments}
This research is supported by the US DOE under 
contract DE-FG02-87ER40331.A008.
\end{acknowledgments}

\newpage

\end{document}